\documentclass[a4paper]{jpconf}
\usepackage{graphicx}
\usepackage{subfig}
\usepackage{tabularx}
\usepackage{amsmath}

\def\la{\mathrel{\hbox{\rlap{\hbox{\lower4pt\hbox{$\sim$}}}\hbox{$<$}}}}
\def\ga{\mathrel{\hbox{\rlap{\hbox{\lower4pt\hbox{$\sim$}}}\hbox{$>$}}}}

\def\fm{\hbox{$.\!\!^{\rm m}$}}

\def\farcs{\hbox{$.\!\!^{\prime\prime}$}}

\begin{document}
\title{Galaxy peculiar velocities in the Zone of Avoidance}

\author{K Said, R C Kraan-Korteweg and T H Jarrett}

\address{Astrophysics, Cosmology and Gravity Centre (ACGC), Department of Astronomy, University
of Cape Town, Private Bag X3, Rondebosch 7701, South Africa.}

\ead{khaled@ast.uct.ac.za, kraan@ast.uct.ac.za, jarrett@ast.uct.ac.za}

\begin{abstract}
Dust extinction and stellar confusion of the Milky Way hinder the detection of galaxies at low Galactic latitude, creating the so-called Zone of Avoidance (ZoA). This has hampered our understanding of the local dynamics, cosmic flow fields and the origin of the Cosmic Microwave Background dipole. The ZoA ($|b| \le 5^\circ$) is also excluded from the ``whole-sky" Two Micron All-Sky Survey (2MASS) Redshift Survey (2MRS) and 2MASS Tully-Fisher Survey (2MTF). The latter aims to provide distances and peculiar velocities for all bright inclined 2MASS galaxies with $K_s^o$  $\leq 11\fm25$. Correspondingly, knowledge about the density distribution in the ZoA remains limited to statistical interpolations. To improve on this bias we pursued two different surveys to fill in the southern and northern ZoA. These data will allow a direct measurement of galaxy peculiar velocities. In this paper we will present a newly derived optimized Tully-Fisher (TF) relation that allow accurate measures of galaxy distances and peculiar velocities for dust-obscured galaxies. We discuss further corrections for magnitudes and biases and present some preliminary results on flow fields in the southern ZoA.
\end{abstract}

\section{Introduction}
Galaxies are not uniformly distributed in space. They instead tend to accumulate in over-dense regions such as clusters, walls and filaments, which surround low-dense regions called voids. To fully understand the dynamics of the Local Group (LG), cosmic flow fields, and the origin of the observed dipole in the Cosmic Microwave Background (CMB), we need to trace this Large-Scale-Structure (LSS) through all-sky surveys, not only with galaxy positions, but also their velocities \cite{2000A&ARv..10..211K,2008MNRAS.386.2221L}. Velocities of galaxies can be decomposed into two components: a recession velocity due to the uniform expansion of the universe and a peculiar velocity caused by density inhomogeneities. The peculiar velocity (its radial component) is measured as the difference between the observed recession velocity of a galaxy and its Hubble velocity. Because they arise as a result of gravity, peculiar velocities can be used as tracers of the matter distribution, both luminous and dark.

Separation of these two components requires a distance indicator. Since the discovery of the empirical Tully-Fisher relation (TF; \cite{1977A&A....54..661T}), it has been widely used as the distance indicator in peculiar velocity surveys. The Two Micron All-Sky (2MASS; \cite{2006AJ....131.1163S}) Tully-Fisher Survey (2MTF; \cite{2008AJ....135.1738M}) intends to provide a complete Tully-Fisher analysis (i.e. distances and peculiar velocities) of all bright inclined spirals in the 2MASS Redshift Survey (2MRS; \cite{2012ApJS..199...26H}). Although the use of a Near InfraRed (NIR) TF analysis will reduce the impact of dust extinction of the Milky Way because it suffers less from extinction, the Zone of Avoidance (ZoA) is not sampled in the 2MTF ($|b|<5^\circ$; $|b|<8^\circ$ for $|l|<30^\circ$). For cosmological applications, different methods have been employed to interpolate across this part of the sky e.g. \cite{1994ApJ...423L..93L,1995MNRAS.275..797K}. This may be valid for the conspicuous features that cross the ZoA, such as the Great Attractor (GA), Perseus-Pisces Supercluster, Puppis Cluster or the Local Void. However, more recent ZoA surveys reveal large-scale structures that were not anticipated. Blind HI surveys in the southern ZoA \cite{2000AJ....119.2686H,2005IAUS..216..203K} have opened an elegant way to fill in this gap. These kind of surveys do not suffer from foreground extinction. We aim to use these data to determine reliable peculiar velocities for the galaxies in the ZoA which are excluded from the 2MRS and the 2MTF. The 2MTF thus supplemented with our data will result, for the first time, in a complete all-sky peculiar velocity survey.

This paper is organized as follows. In Section 2 we discuss the data from both the southern and northern ZoA. Section 3 provides a recipe on how to derive peculiar velocities for field galaxies in the ZoA. A preliminary flow field for a subsample of the southern ZoA is presented in Section 4.

\section{Data in the ZoA}
The extent of the Zone of Avoidance is different in various wavebands. While in the optical and the NIR the major problems are dust extinction and stellar confusion, systematic HI surveys do not suffer from these effects. Therefore, two independent observing programmes have been started to obtain HI observations for galaxies in both the southern and the northern ZoA. 

\subsection{HI Parkes Deep ZoA Survey}
The HI Parkes Deep Zone of Avoidance Survey (HIZoA) is a blind HI survey of the southern ZoA ($|b| \le 5^\circ$; Dec $< +15^\circ$), conducted on the 64 m Parkes radio telescope \cite{2005RvMA...18...48K,2005AJ....129..220D,2008MSC}. The survey covers the whole southern ZoA visible from Parkes out to 12700 km/s, and resulted in about 900 detected galaxies with rms~$\sim 6$mJy/beam. We have performed an imaging NIR $J$, $H$, $K_s$ bands follow-up survey of all HIZoA-galaxies with the InfraRed Survey Facilty (IRSF) on the 1.4m telescope at Sutherland, South Africa \cite{2011arXiv1107.1096W}. 

\subsection{2MASX Redshift ZoA} 
In 2009, Kraan-Korteweg and collaborators started an observing programme to obtain HI redshifts for all bright galaxies in the 2MASS Extended Source Catalog (2MASX) \cite{2000AJ....119.2498J} with $K_s^o < 11.25$ mag in the ZoA ($|b| < 10^\circ$) which do not have any previous redshift determinations. The Nan\c{c}ay Radio Telescope (NRT) in France and the Parkes radio telescope in Australia have been used to cover the entire ZoA. Over a thousand galaxies without previous redshifts have been observed with the NRT (Dec $ > -38^\circ$), resulting in more than 250 detected galaxies \cite{2012MSC}. Parkes observations started in 2013 and 74 galaxies without previous redshifts have been observed to date.

\section{Methodology}
The Tully-Fisher relation is a correlation between rotational velocities and absolute magnitudes of spiral galaxies. The rotational velocity is extracted directly from the HI spectral data, and the apparent magnitude is measured from the imaging data. Deriving peculiar velocities from the TF relation, requires the calibration of a global TF relation from a sample of galaxies with known distances. The peculiar velocity is defined as the difference in the Hubble velocity according to the TF distance and the observed velocity from its redshift. 

\subsection{Tully-Fisher template relation in $J$, $H$, and $K_s$ bands}
The TF calibration sample should be unbiased. Although the galaxies of the template sample lie far from the Galactic Plane, their photometric and spectroscopic data should be treated in the same way as any peculiar velocity sample to which it will be applied. The currently most elaborate and extensive NIR TF calibration in $J$, $H$, and $K_s$ bands \cite{2008AJ....135.1738M} of 888 galaxies is based on the total magnitude as the closest proxy to total mass. However, 2MASS does not trace the lower surface brightness features very well. This is particularly severe in areas of high dust extinction and stellar density. We therefore prefer to work with isophotal magnitudes which should not be affected by those effects. 

To demonstrate this, we construct a comparison between two different surveys, the 2MASX \cite{2000AJ....119.2498J} and the IRSF galaxy Catalog \cite{2011arXiv1107.1096W}. The depth of the IRSF survey (10 min exposure time) and four times higher spatial resolution (0\farcs 45/pix) results in a 2 mag deeper survey compared to 2MASX. We select a subsample of 66 galaxies, observed in both 2MASX and the IRSF Catalog and compare their total extrapolated magnitudes and isophotal magnitudes (measured in an elliptical aperture defined at the $K_s=20$ mag arcsec$^{-2}$). 

Figure 1 displays the offsets between 2MASX and IRSF, $\Delta m = m(\text{2MASS})-m(\text{IRSF})$, for both total and isophotal magnitude in the $J$, $H$, and $K_s$ bands. The deeper observations with high optical resolution prove that the 2MASX total magnitudes have large scatter and are on average underestimated. Note that such systematic offsets produce artificial flow in the analysis of peculiar velocities. This effect completely disappears when we use isophotal magnitudes.

This result confirms that it is imperative to use the isophotal magnitude for $|b| < 5^\circ$, or any other region of high dust obscuration. Using isophotal magnitudes on the other hand can equally well be applied in dust and star-free areas of the sky, hence allow real whole-sky coverage.

We therefore re-calibrated the NIR $J$, $H$, $K_s$ bands TF relation following the exact same steps and procedures as outlined in \cite{2008AJ....135.1738M}, but for isophotal magnitudes and including Galactic extinction correction according to \cite{2010MNRAS.401..924R}.

\begin{figure}[h]
\begin{center}
\includegraphics[width=12.5cm,height=14.5cm]{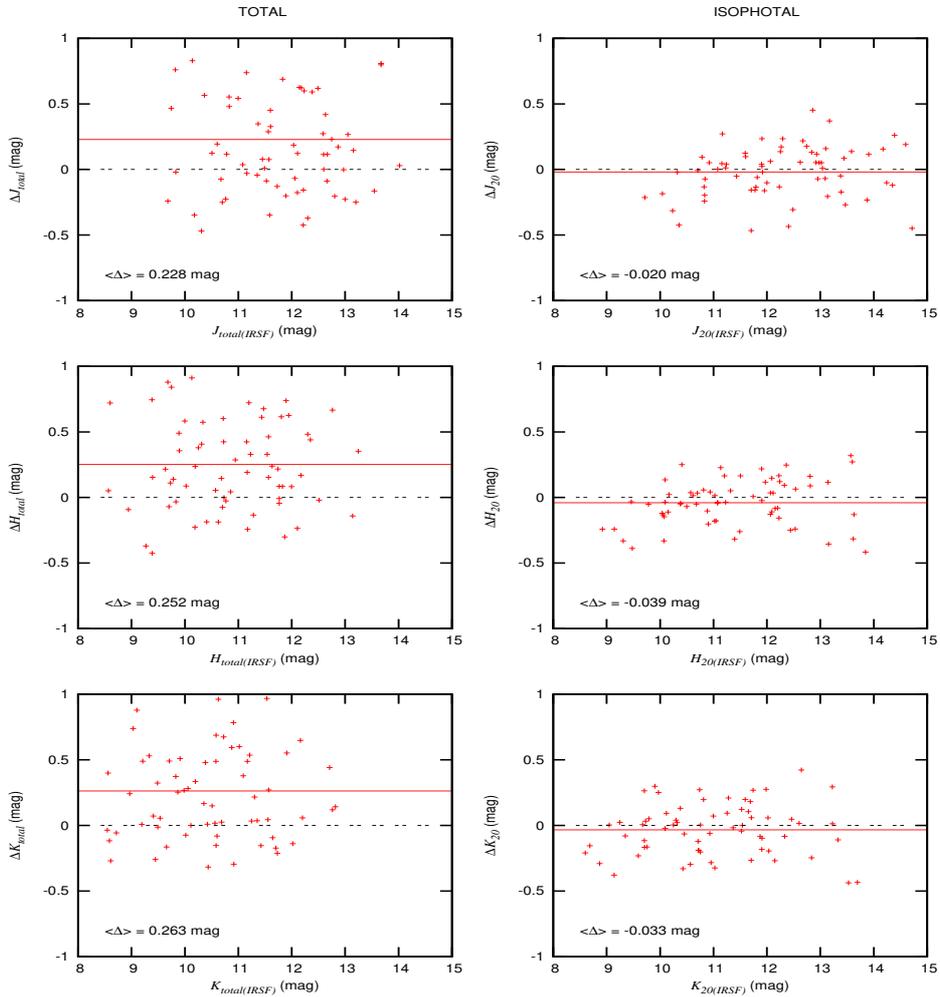}
\caption{Comparison of 2MASS and IRSF photometry of ZoA galaxies. The left panels display the total magnitudes and the right panels display the isophotal magnitudes with $J$, $H$, and $K_s$ band arranged from top to bottom. In all panels, solid red lines mark the mean of the difference, and the dashed black lines is the zero-line. The mean difference is given in the bottom left corner of each plot.}
\end{center}
\end{figure} 

\subsection{Peculiar velocity in the ZoA}
To test our newly calibrated TF relation we used a preliminary sample in the ZoA based on a subsample of HIZoA galaxies with $v_{hel} = 6000$ km/s that have bona-fide counterparts in the IRSF catalogue. We consider here only 120 galaxies with high-fidelity photometric and spectroscopic parameters, after applying additional restrictions on inclination ($i$) and linewidth ($W$). All photometric and spectroscopic data were corrected to be consistent with the TF template relation. We confirm that the isophotal magnitudes lead to a significant decrease in the TF scatter and reduce the overall offset (systematic flow) compared to total magnitudes.

\section{Early results}
The majority of the derived peculiar velocities for the 120 bright, inclined spiral galaxies lie within $|v_{pec}| \leq 1000$ km/s, with the largest outliers out to $|v_{pec}| \leq 2500$ km/s. The latter seem to originate from fairly face-on or low mass galaxies. We used these peculiar velocities to construct a flow-field map in the ZoA (Figure 2). This map is obtained by interpolating the peculiar velocities across the Galactic longitude and the recession velocity.  Each dot in the map is a galaxy colour-coded by its measured peculiar velocity. The black star presents the centre of the GA \cite{1999A&A...352...39W}. The map clearly shows the infall into the GA. 

The selection of galaxies with $v_{obs}\leq 6000$~km/s limits our ability to quantify the backside infall toward the GA, and  the infall into the much more remote Shapley Concentration (SC). To improve on the GA infall signal we started to extend our sample to $v_{obs}\leq 9000$~km/s in 2012. For even larger redshifts the sample becomes very sparse and the errors on measured peculiar velocities become comparable to the peculiar velocities themselves. The peculiar velocity error is calculated by combining the absolute magnitude error, velocity width error, and the intrinsic scatter in TF-template relation in a quadratic form. Most of the galaxies have errors of a few hundred km/s, which is a typical value for most methods available today. 

\begin{figure}[h]
\begin{center}
\includegraphics[scale=0.55]{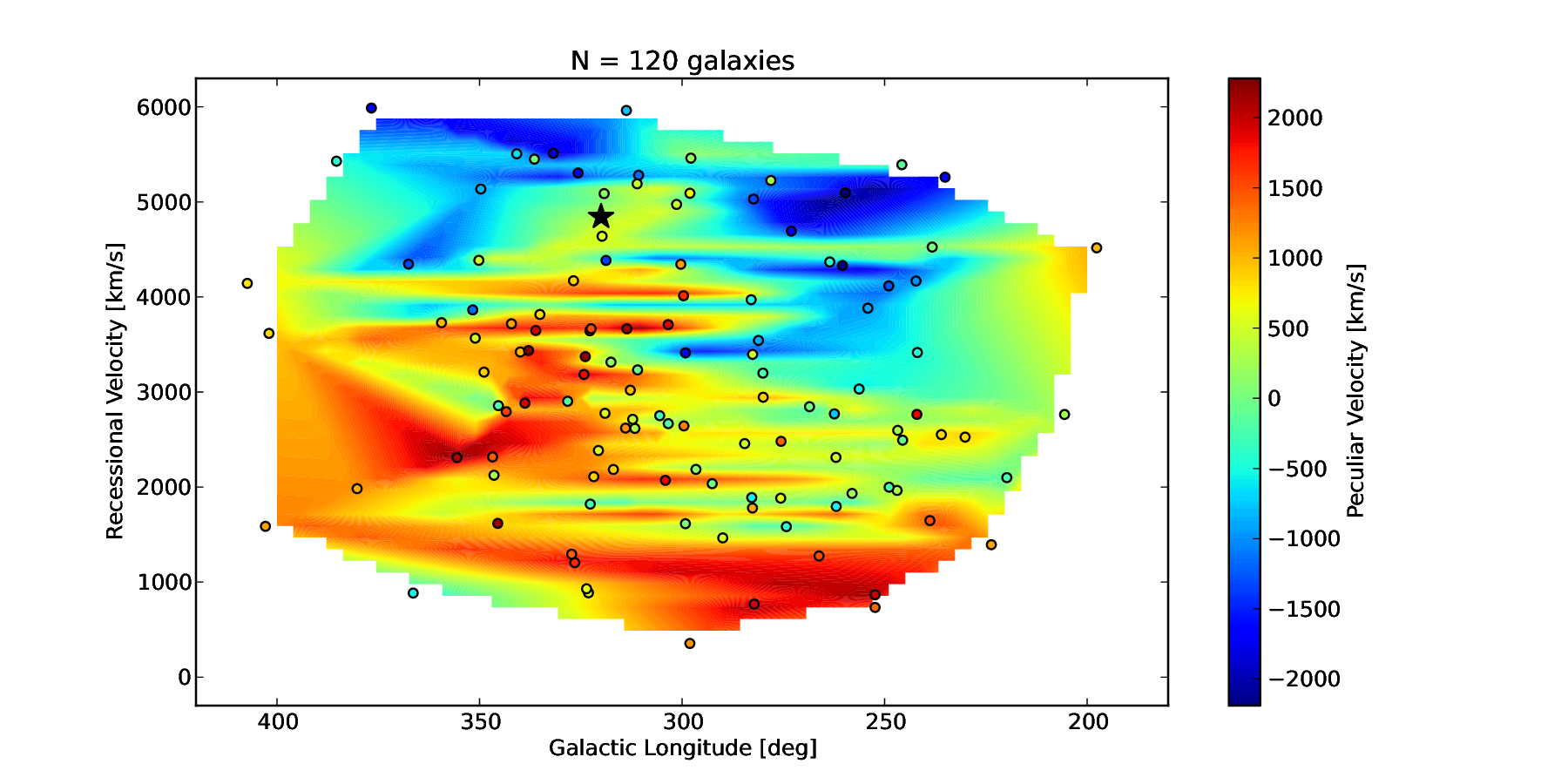}
\caption{Peculiar velocity map of galaxies with recession velocities below $6000$~km/s and $-2500$~km/s $< v_{pec} < 2500 $~km/s, obtained by interpolating the peculiar velocities across the Galactic longitude and recession velocity. All galaxies with $\log W > 1.8$ are used here. Individual galaxies are colour-coded by their derived peculiar velocities (in km/s). The centre of the GA is plotted as a large black star.}
\end{center}
\end{figure} 

For a first interpretation of the resulting data, we compare our peculiar velocities with a GA model proposed by Faber \& Burstein \cite{1988lsmu.book..115F}. The model describes the influence of the GA $v_{pec,A}=v_A(r_A/d_A)[(d_A^2+c_A^2)/(r_A^2+c_A^2)]^{(n_A+1)/2},$ where $v_{pec}$ is the predicted peculiar velocity, $r_A$ is the distance of a galaxy from the infall centre, $d_A$ is the distance of the infall centre from the Local Group, $v_A$ is the peculiar velocity generated by the GA at the position of the LG, and $c_A$ is the core radius of the centre of the GA. We adopted the values of parameters given in \cite{1990RPPh...53..421B}, $n_A=1.7$, $v_A=535$~km/s and $c_A=1430$~km/s. For $d_A$ we used 4844~km/s, following \cite{1999A&A...352...39W}. To compare the single GA model to our data, we restricted our sample to the most accurate subsample using only the galaxies with $\log W > 2.3$ and $i<0.5$ in the GA region ($290^\circ<l<350^\circ$; $|b| \leq 5^\circ$). Figure 3 depicts the data with the GA model. The shaded area presents our observational limit of $v_{obs} \leq 6000$ km/s. While the errors of the peculiar velocities increase with distance (as expected), the peculiar velocities seem to follow the shape of the early GA model quite well (although no quantitative analysis has been done as yet).

\begin{figure}[h]
\includegraphics[width=24pc]{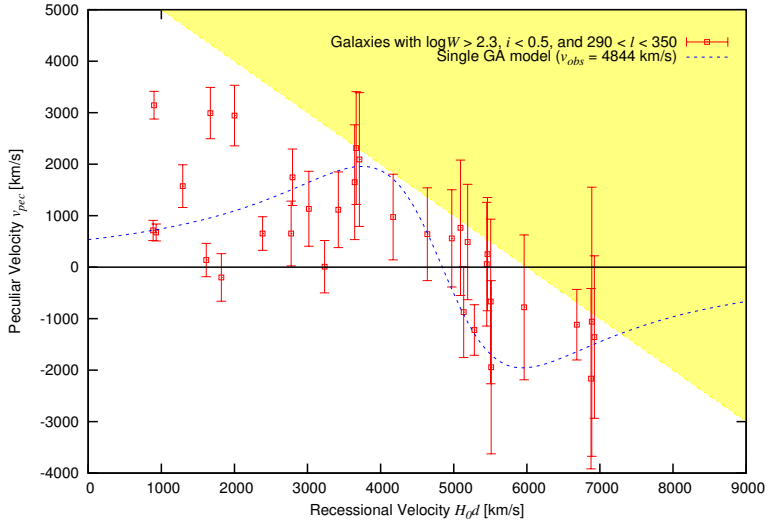}\hspace{2pc}%
\begin{minipage}[b]{14pc}\caption{\label{label}Peculiar velocity as a function of recessional velocity for a subsample of galaxies with $\log W > 2.3$, $i<0.5$ in the region $290^\circ<l<350^\circ$. The dotted curve presents the GA model. The error bars reflect the peculiar velocity errors. The deviation of some points from the curve probably due to selection effects.}
\end{minipage}
\end{figure}

\section{Conclusions}
We have shown that a peculiar velocity analysis in the ZoA is feasible if a NIR isophotal TF relation is applied. This new tool and new data provide a powerful tool to trace the matter distribution in the ZoA. Our preliminary (yet incomplete) data with $v_{hel} \leq 6000$ km/s show clear infall into the GA. The data set has since been extended to include all the HI galaxies in the southern ZoA within $v_{obs} \leq 9000$ km/s. This will enable us to test multi-attractor models, such as GA plus the Shapley Concentration. Combining our data with that from 2MTF will lead to a complete all-sky peculiar velocity survey. 

\section*{Acknowledgements}
This work is based upon research supported by the South African National Research Foundation and Department of Science and Technology. The authors thank Wendy Williams for communicating her data in the ZoA. KS is very grateful to Dr. Maciej Bilicki for his reading and comments on the paper. We acknowledge the HIZOA survey team for early access to the data. This publication makes use of data products from the Two Micron All Sky Survey, which is a joint project of the University of Massachusetts and the Infrared Processing and Analysis Center, funded by the National Aeronautics and Space Administration and the National Science Foundation.

\end{document}